\documentclass{aa}
\usepackage{graphicx}
\usepackage{multirow}
\usepackage{subcaption}
\usepackage{amssymb}
\usepackage{tabularx}
\usepackage{booktabs}
\usepackage[usenames,dvipsnames]{xcolor}
\usepackage[labelfont=bf]{caption}
\usepackage{txfonts}
\usepackage[angle-symbol-over-decimal=true,
            uncertainty-mode = separate,
            separate-uncertainty-units = single,
            bracket-ambiguous-numbers = false,
            detect-weight]{siunitx}
\usepackage[hidelinks]{hyperref}
\usepackage{orcidlink}
\usepackage[capitalize]{cleveref}
\crefname{section}{Sect.}{Sects.}
\Crefname{section}{Section}{Sections}
\crefname{figure}{Fig.}{Figs.}
\Crefname{figure}{Figure}{Figures}
\crefname{equation}{Eq.}{Eqs.}
\Crefname{equation}{Equation}{Equations}
\crefname{table}{Table}{Tables}
\Crefname{table}{Table}{Tables}
\creflabelformat{equation}{#2(#1)#3}

\newcommand{\Om}{\Omega_{\mathrm{m}}}

\newcommand{\papercatalog}{I}       
\newcommand{\paperbmodes}{II}       
\newcommand{\paperconfig}{III}      
\newcommand{\paperharmonic}{IV}     
\newcommand{\papersims}{V}          

\hypersetup{
    colorlinks=true,
    linkcolor=MidnightBlue,
    filecolor=MidnightBlue,
    citecolor=MidnightBlue,
    urlcolor=MidnightBlue,
}

\idline{UNIONS-3500 WL Paper II}

\begin{document}



\newcommand{\ebthetaXipMin}{\num{12}}
\newcommand{\ebthetaXipMax}{\num{83}}
\newcommand{\ebthetaXimMin}{\num{12}}
\newcommand{\ebthetaXimMax}{\num{83}}

\newcommand{\ebcovCondE}{\num{1.5e+05}}
\newcommand{\ebcovCondB}{\num{2.0e+05}}
\newcommand{\ebcovCondAmb}{\num{1.8e+10}}
\newcommand{\ebcovCondFull}{\num{2.7e+10}}
\newcommand{\ebcovNbins}{\num{20}}


\newcommand{\configPteSixThreeXip}{\num{0.31}}
\newcommand{\configPteSixThreeXim}{\num{0.26}}
\newcommand{\configPteSixThreeCombined}{\num{0.18}}
\newcommand{\configPteSixThreeCosebis}{\num{0.78}}
\newcommand{\configPteSixThreeCosebisTwenty}{\num{0.94}}
\newcommand{\configPteSixThreeXipFull}{\num{0.45}}
\newcommand{\configPteSixThreeXimFull}{\num{0.16}}
\newcommand{\configPteSixThreeCombinedFull}{\num{0.40}}
\newcommand{\configPteSixThreeCosebisFull}{\num{1.37e-05}}
\newcommand{\configPteSixThreeCosebisTwentyFull}{\num{1.08e-04}}
\newcommand{\configPteFiveXip}{\num{0.51}}
\newcommand{\configPteFiveXim}{\num{0.005}}
\newcommand{\configPteFiveCombined}{\num{0.030}}
\newcommand{\configPteFiveCosebis}{\num{0.94}}
\newcommand{\configPteFiveCosebisTwenty}{\num{0.61}}
\newcommand{\configPteFiveXipFull}{\num{0.026}}
\newcommand{\configPteFiveXimFull}{\num{7.71e-04}}
\newcommand{\configPteFiveCombinedFull}{\num{0.004}}
\newcommand{\configPteFiveCosebisFull}{\num{7.37e-10}}
\newcommand{\configPteFiveCosebisTwentyFull}{\num{4.15e-07}}
\newcommand{\configPteEightXip}{\num{0.020}}
\newcommand{\configPteEightXim}{\num{0.047}}
\newcommand{\configPteEightCombined}{\num{0.014}}
\newcommand{\configPteEightCosebis}{\num{0.60}}
\newcommand{\configPteEightCosebisTwenty}{\num{0.81}}
\newcommand{\configPteEightXipFull}{\num{0.039}}
\newcommand{\configPteEightXimFull}{\num{0.020}}
\newcommand{\configPteEightCombinedFull}{\num{0.018}}
\newcommand{\configPteEightCosebisFull}{\num{4.94e-06}}
\newcommand{\configPteEightCosebisTwentyFull}{\num{2.50e-05}}
\newcommand{\configPteElevenThreeXip}{\num{0.78}}
\newcommand{\configPteElevenThreeXim}{\num{9.40e-04}}
\newcommand{\configPteElevenThreeCombined}{\num{0.014}}
\newcommand{\configPteElevenThreeCosebis}{\num{0.82}}
\newcommand{\configPteElevenThreeCosebisTwenty}{\num{0.83}}
\newcommand{\configPteElevenThreeXipFull}{\num{0.10}}
\newcommand{\configPteElevenThreeXimFull}{\num{2.94e-06}}
\newcommand{\configPteElevenThreeCombinedFull}{\num{9.07e-04}}
\newcommand{\configPteElevenThreeCosebisFull}{\num{1.70e-10}}
\newcommand{\configPteElevenThreeCosebisTwentyFull}{\num{1.16e-08}}


\newcommand{\clPteFiveFid}{\num{0.003}}
\newcommand{\clPteFiveFull}{\num{2.53e-04}}
\newcommand{\clPteSixThreeFid}{\num{0.30}}
\newcommand{\clPteSixThreeFull}{\num{0.13}}
\newcommand{\clPteEightFid}{\num{0.43}}
\newcommand{\clPteEightFull}{\num{0.11}}
\newcommand{\clPteElevenThreeFid}{\num{5.74e-10}}
\newcommand{\clPteElevenThreeFull}{\num{1.78e-12}}


\newcommand{\harmCosebisPteSixThreeFull}{\num{1.61e-05}}
\newcommand{\harmCosebisChisqSixThreeFull}{\num{32.03}}
\newcommand{\harmCosebisPteElevenThreeFull}{\num{4.64e-18}}
\newcommand{\harmCosebisChisqElevenThreeFull}{\num{93.92}}
\newcommand{\harmCosebisPteFiveFull}{\num{6.41e-09}}
\newcommand{\harmCosebisChisqFiveFull}{\num{49.33}}
\newcommand{\harmCosebisPteEightFull}{\num{4.72e-05}}
\newcommand{\harmCosebisChisqEightFull}{\num{29.58}}
\newcommand{\cfgCosebisPteSixThreeFull}{\num{1.37e-05}}
\newcommand{\cfgCosebisChisqSixThreeFull}{\num{32.40}}
\newcommand{\cfgCosebisPteElevenThreeFull}{\num{1.70e-10}}
\newcommand{\cfgCosebisChisqElevenThreeFull}{\num{57.15}}
\newcommand{\cfgCosebisPteFiveFull}{\num{7.37e-10}}
\newcommand{\cfgCosebisChisqFiveFull}{\num{54.00}}
\newcommand{\cfgCosebisPteEightFull}{\num{4.94e-06}}
\newcommand{\cfgCosebisChisqEightFull}{\num{34.69}}


\newcommand{\harmCosebisPteSixThreeFid}{\num{0.60}}
\newcommand{\harmCosebisChisqSixThreeFid}{\num{4.59}}
\newcommand{\harmCosebisPteElevenThreeFid}{\num{0.85}}
\newcommand{\harmCosebisChisqElevenThreeFid}{\num{2.63}}
\newcommand{\harmCosebisPteFiveFid}{\num{0.73}}
\newcommand{\harmCosebisChisqFiveFid}{\num{3.59}}
\newcommand{\harmCosebisPteEightFid}{\num{0.40}}
\newcommand{\harmCosebisChisqEightFid}{\num{6.17}}
\newcommand{\cfgCosebisPteSixThreeFid}{\num{0.78}}
\newcommand{\cfgCosebisChisqSixThreeFid}{\num{3.25}}
\newcommand{\cfgCosebisPteElevenThreeFid}{\num{0.82}}
\newcommand{\cfgCosebisChisqElevenThreeFid}{\num{2.88}}
\newcommand{\cfgCosebisPteFiveFid}{\num{0.94}}
\newcommand{\cfgCosebisChisqFiveFid}{\num{1.79}}
\newcommand{\cfgCosebisPteEightFid}{\num{0.60}}
\newcommand{\cfgCosebisChisqEightFid}{\num{4.54}}

   \title{UNIONS-3500 Weak Lensing: \paperbmodes.\\$B$-mode validation for cosmic shear}

   \author{C.~Daley\orcidlink{0000-0002-3760-2086}
          \inst{1}
          \thanks{\email{cail.daley@cea.fr}}
          \and
          A.~Guinot\orcidlink{0000-0002-5068-7918}
          \inst{2}
          \and
          S.~Guerrini\orcidlink{0009-0004-3655-4870}
          \inst{3}
          \and
          F.~Hervas-Peters\orcidlink{0009-0008-1839-2969}
          \inst{1}
          \and
          L.~W.~K.~Goh\orcidlink{0000-0002-0104-8132}
          \inst{4,5}
          \and
          C.~Murray\orcidlink{0000-0002-4668-1273}
          \inst{1}
          \and
          M.~Kilbinger\orcidlink{0000-0001-9513-7138}
          \inst{1}
          \and
          A.~Wittje\orcidlink{0000-0002-8173-3438}
          \inst{6}
          \and
          H.~Hildebrandt\orcidlink{0000-0002-9814-3338}
          \inst{6}
          \and
          M.~J.~Hudson\orcidlink{0000-0002-1437-3786}
          \inst{7,8,9}
          \and
          L.~van~Waerbeke\orcidlink{0000-0002-2637-8728}
          \inst{10}
          \and
          A.~W.~McConnachie\orcidlink{0000-0003-4666-6564}
          \inst{11}
          }

   \institute{%
        Universit\'e Paris-Saclay, Universit\'e Paris Cit\'e, CEA, CNRS, AIM, F-91191 Gif-sur-Yvette, France
    \and
        Department of Physics, McWilliams Center for Cosmology and Astrophysics, Carnegie Mellon University, Pittsburgh, PA 15213, USA
    \and
        Universit\'e Paris Cit\'e, Universit\'e Paris-Saclay, CEA, CNRS, AIM, F-91191 Gif-sur-Yvette, France
    \and
        Institute for Astronomy, University of Edinburgh, Royal Observatory, Blackford Hill, Edinburgh EH9 3HJ, UK
    \and
        Higgs Centre for Theoretical Physics, School of Physics and Astronomy, The University of Edinburgh, Edinburgh EH9 3FD, UK
    \and
        Ruhr University Bochum, Faculty of Physics and Astronomy, Astronomical Institute (AIRUB), German Centre for Cosmological Lensing, 44780 Bochum, Germany
    \and
        Department of Physics and Astronomy, University of Waterloo, 200 University Avenue West, Waterloo, Ontario N2L 3G1, Canada
    \and
        Waterloo Centre for Astrophysics, University of Waterloo, Waterloo, Ontario N2L 3G1, Canada
    \and
        Perimeter Institute for Theoretical Physics, 31 Caroline St. North, Waterloo, ON N2L 2Y5, Canada
    \and
        Department of Physics and Astronomy, University of British Columbia, 6224 Agricultural Road, V6T 1Z1, Vancouver, Canada
    \and
        NRC Herzberg Astronomy and Astrophysics, 5071 West Saanich Road, Victoria, BC V8Z 6M7, Canada
    }

   \date{Received XXXX; accepted YYYY}

  \abstract{%
At Stage-III sensitivities, cosmic shear $B$ modes unambiguously indicate systematic contamination and are often used to inform data selection and scale cuts for cosmological inference.
We validate $B$ modes for the Ultraviolet Near-Infrared Optical Northern Survey (UNIONS)-3500 (\SI{2894}{\square\deg}, $n_\mathrm{eff} \approx \SI{5.0}{arcmin\tothe{-2}}$) using three $E$/$B$-separable statistics: pure-mode correlation functions $\xi_\pm^{\mathrm{B}}(\theta)$, Complete Orthogonal Sets of $E$/$B$-mode Integrals (COSEBI) $B$-mode amplitudes $B_n$, and harmonic-space power spectra $C_\ell^{BB}$.
For each statistic, we compute probability-to-exceed (PTE) values over a two-dimensional grid of scale-cut boundaries; our adopted cuts lie in broad stable regions of acceptable PTE.
$B$-mode detections and PTE failures on initial catalog versions led us to investigate galaxy size cuts and stellar halo masking.
After cuts, all three statistics pass the null test (minimum PTE $= \configPteSixThreeCombined$).
Before scale cuts, we measure an oscillatory COSEBI $B$-mode pattern consistent with repeating additive shear bias, a detector-level effect seen across multiple Stage-III surveys including CFHTLenS, which used the same MegaCam camera; scale cuts that exclude the charge-coupled device (CCD) angular scale suppress it.
Although these statistics probe the same two-point shear field, scale cuts in one do not map exactly onto cuts in another, because their respective filter functions weight angular scales differently.
The most conservative validation therefore requires scale and sample selections that pass null tests across all frameworks simultaneously, an approach that applies directly to Stage-IV surveys where systematic errors dominate.}


   \maketitle
%

\section{Introduction}

Weak gravitational lensing distorts the observed shapes of distant galaxies, encoding information about the intervening matter distribution along the line of sight \citep{blandford.etal91, miraldaescude91, kaiser92}.
First detections came in 2000 \citep{bacon.etal00, kaiser.etal00, vanwaerbeke.etal00, wittman.etal00}.
The scalar gravitational potential produces a shear field that is curl-free to leading order and can be decomposed into gradient-like $E$ modes and curl-like $B$ modes; under the Born approximation, scalar density perturbations source only $E$-mode power \citep{stebbins96, kamionkowski.etal97}.
Higher-order effects (lens-lens coupling, source clustering, and intrinsic alignments) can produce $B$ modes \citep{hilbert.etal09, schneider.etal02b, crittenden.etal02}, but these are orders of magnitude below current sensitivity \citep{cooray.hu02}.
They are likely only accessible to future ultra-high-precision measurements, potentially through cross-correlation techniques involving CMB lensing rotation reconstructions and large-scale-structure tracers \citep{robertson.etal25}.
For galaxy shear measurements at current sensitivity, detected $B$-mode power indicates residual observational systematic effects rather than cosmological signal, though the absence of $B$ modes does not rule out systematic contamination.

Stage-III cosmic shear surveys have measured the structure growth parameter $S_8 \equiv \sigma_8 \sqrt{\Om/0.3}$ at percent-level precision, with some reporting up to ${\sim}3\,\sigma$ tension with \textit{Planck} \citep{asgari.etal21, abbott.etal22, li.etal23, planck20}.
The Kilo-Degree Survey (KiDS)-Legacy \citep{wright.etal25} found $S_8 = 0.815^{+0.016}_{-0.021}$ (\num{0.73}\,$\sigma$ from \textit{Planck}) and the Dark Energy Survey (DES) Y6 \citep{abbott.etal26} found $S_8 = 0.798^{+0.014}_{-0.015}$ ($1.1\,\sigma$); both achieved $2\%$ precision on $S_8$, comparable to CMB constraints from \textit{Planck} alone.
Stage-IV surveys will reduce statistical uncertainties further, shifting the error budget toward systematic effects; current data offer the opportunity to validate these methods before that transition.

Several methods separate $E$- and $B$-mode contributions in shear two-point statistics.
Aperture mass dispersion \citep{schneider96, schneider.etal98} is the foundational real-space method for $E$/$B$ separation.
In practice, however, its evaluation requires knowledge of the shear correlation functions down to zero separation, which is unavailable on a finite observed interval and leads to $E$/$B$ mixing \citep{schneider.eifler.krause10, schneider.etal22}.
Complete Orthogonal Sets of $E$/$B$-mode Integrals (COSEBIs; \citealt{schneider.eifler.krause10, asgari.schneider.simon12}) generalize this approach, compressing all $E$/$B$-separable information from $\xi_\pm(\theta)$ on a finite angular interval into discrete orthogonal modes.
A small number of modes captures the cosmological signal; by construction, the COSEBI basis excludes the ambiguous modes that cannot be assigned pure $E$ or $B$ character on a finite interval.
Pure-mode correlation functions $\xi_\pm^{\mathrm{E/B}}(\theta)$ \citep{schneider.etal22} are an alternative real-space representation that decomposes $\xi_\pm$ into $E$-mode, $B$-mode, and ambiguous components through integral transforms derived from the COSEBI basis functions.
Harmonic-space estimators measure $C_\ell^{EE}$ and $C_\ell^{BB}$ directly but require careful treatment of mask-induced $E$/$B$ mixing; catalog-based methods that evaluate spherical harmonic transforms at source positions \citep{wolz.etal25} avoid the pixelization artifacts that complicate standard implementations.
The \textsc{HybridEB} Fourier band-power estimator \citep{becker.rozo16} constructs $\ell$-space band-powers from linear combinations of binned $\xi_\pm$, projecting out ambiguous modes to minimize $E$/$B$ mixing.
DES Y6 adopted this estimator \citep{abbott.etal26}, although COSEBIs are more widely used in the Stage-III literature, where they have traced systematic signatures across multiple surveys \citep{asgari.etal19a}.
These statistics represent $B$-mode power in complementary bases, and the mapping between them is not one-to-one, so contamination excluded by scale cuts in one basis may remain visible in another.
Comparing across bases can expose systematics that any single framework would absorb.

Stage-III analyses have found $B$-mode contamination from a range of systematic effects.
COSEBI analyses of the Canada-France-Hawaii Telescope Lensing Survey (CFHTLenS), KiDS-450, and DES-SV detected significant $B$ modes at 2--5\,$\sigma$, linked to repeating additive shear bias, PSF leakage, and photometric selection effects \citep{asgari.etal19a}.
Repeating additive shear biases can arise from detector-level defects that imprint a constant ellipticity offset on each CCD, varying across the focal plane; because this pattern is fixed in instrument coordinates, it repeats with every pointing.
The bias is constant within each CCD but jumps at chip boundaries; these discontinuities carry curl, generating $B$-mode power at CCD angular scales.
Hyper Suprime-Cam (HSC) Y3 used harmonic-space power spectra and found significant large-scale $B$ modes \citep{li.etal23, dalal.etal23}. A dedicated PSF-systematics study identified PSF fourth-moment leakage as a leading additive contaminant in the affected region \citep{zhang.etal23}.
KiDS-Legacy tested with COSEBIs, traced an initial failure to astrometric systematics, and applied conservative masking with negligible impact on $S_8$ \citep{wright.etal25}.
DES Y3 reported null detections using pseudo-$C_\ell$ and COSEBIs \citep{gatti.etal21}.
However, \citet{jefferson.etal25} applied the \textsc{HybridEB} estimator uniformly to DES-Y3 and HSC-Y3 data and found $B$-mode failures in specific tomographic bins not seen with the estimators originally used.
DES Y6 subsequently adopted the \textsc{HybridEB} estimator and passed \citep{abbott.etal26}.
Because each estimator weights angular scales differently and deprojects ambiguous modes differently, the same data can pass one null test and fail another.
The most informative validation requires consistent passage across multiple statistics.

The Ultraviolet Near-Infrared Optical Northern Survey (UNIONS; \citealt{gwyn.etal25}) is surveying \SI{6250}{\square\deg} of northern sky with multi-band imaging from telescopes in Hawai'i:
the Canada-France Imaging Survey (CFIS) contributes $u$- and $r$-band imaging from CFHT,
Pan-STARRS contributes $i$- and $z$-band data,
and Subaru adds $z$-band imaging through the Wide Imaging with Subaru HSC of the \textit{Euclid} Sky (WISHES) and $g$-band imaging through the Waterloo-Hawai'i Institute for Astronomy $g$-band Survey (WHIGS).
Nearly all Stage-III cosmic shear survey area is in the southern sky; UNIONS provides the first deep, wide-area shape catalog in the north, with $r$-band median seeing of $0\farcs7$, enabling high-quality shape measurements.
\citet{guinot.etal22} presented the first UNIONS weak-lensing analysis over \SI{1700}{\square\deg} using the \texttt{ShapePipe} pipeline \citep{farrens.etal22};
the present analysis expands this to \SI{3500}{\square\deg} with an updated catalog (\cref{sec:data}).

We present $B$-mode validation tests for the first UNIONS-3500 cosmic shear cosmology analysis, using pure-mode correlation functions $\xi_\pm^{\mathrm{E/B}}(\theta)$, COSEBIs, and harmonic-space power spectra $C_\ell^{BB}$ to define scale cuts in both angular and multipole space within a blinded analysis framework.
This paper is one of five coordinated UNIONS-3500 weak-lensing publications:
catalog construction (Paper~\papercatalog; \citealt{hervaspaters.etal26a});
the present $B$-mode validation (Paper~\paperbmodes);
configuration-space cosmological constraints (Paper~\paperconfig; \citealt{goh.etal26});
harmonic-space cosmological constraints (Paper~\paperharmonic; \citealt{guerrini.etal26});
and image simulations (Paper~\papersims; \citealt{hervaspaters.etal26b}).
We validate the fiducial scale cuts adopted for cosmological inference in Papers~\paperconfig{} and~\paperharmonic.
\Cref{sec:data} describes the UNIONS shear catalogs and their evolution.
\Cref{sec:methods} details the three $E$/$B$-separable statistics and covariance framework.
\Cref{sec:results} presents measurements and PTE assessments.
\Cref{sec:discussion} discusses systematic trends and methodological implications.

\section{Data}
\label{sec:data}

UNIONS is the largest deep weak-lensing survey of the northern sky, with $r$-band depth of $r \approx 24.2$ at $10\,\sigma$ \citep{gwyn.etal25}.
We analyze four variants of the \texttt{ShapePipe} v1.4 catalog (Paper~\papercatalog, Table~H.1).
The initial catalog uses a loose size cut ($r_{\mathrm{h, gal}}/r_{\mathrm{h, psf}} > 0.5$).
For the size-cut catalog, we tighten this threshold to $r_{\mathrm{h, gal}}/r_{\mathrm{h, psf}} > 0.707$, removing galaxies where shape measurement and leakage correction are least reliable.
For the masked catalog, we retain the tightened size cut and add bright-star ($r < 12$) and faint-star ($12 < r < 18$) halo masking, reducing sky coverage by 13\%.
The relaxed-flags catalog uses the same sample selection as the size-cut catalog but relaxes the \texttt{SExtractor} flags criterion to include deblended objects, increasing the effective galaxy density by 23\% (Paper~\papercatalog).
We introduced these refinements after blinded validation tests showed early signs of systematic contamination (Paper~\papercatalog).
We adopt the size-cut catalog as the fiducial for Papers~\paperbmodes, \paperconfig, and~\paperharmonic.

For each catalog version, we produce shear measurements both with and without the object-wise PSF-leakage correction described in Paper~\papercatalog{}, which models PSF contamination via binned regression in signal-to-noise and galaxy-PSF size ratio; \citet{guerrini.etal25} provide the broader PSF-leakage diagnostic framework used to assess its impact.
We report results from the leakage-corrected catalogs, which are used for cosmological inference in Papers~\paperconfig{} and~\paperharmonic.



\section{Methods}
\label{sec:methods}

\begin{figure*}[t]
\centering
\includegraphics[width=\linewidth]{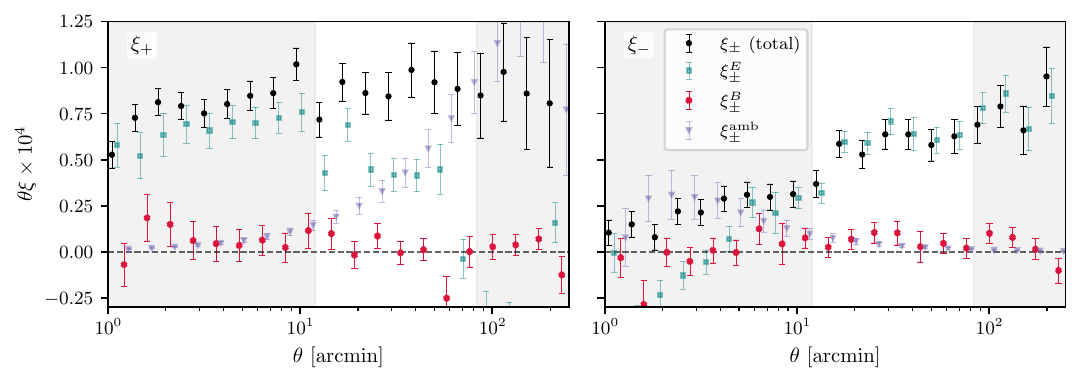}
\caption{Pure $E$/$B$-mode decomposition of the measured shear correlation functions for the fiducial catalog.
Panels show $\xi_+$ (left) and $\xi_-$ (right) decomposed into total (black circles), $E$-mode (teal squares), $B$-mode (crimson $\times$), and ambiguous (purple triangles) components following \citet{schneider.etal22}.
The $E$ modes trace the lensing signal, while $B$ modes---which should vanish for pure gravitational lensing---diagnose systematic contamination.
Ambiguous modes capture boundary effects from the finite angular range.
Shaded regions indicate fiducial scale cuts.
Error bars are derived from semi-analytical covariance propagation (\cref{sec:covariance}).}
\label{fig:pure_eb_decomposition}
\end{figure*}

\subsection{Two-point correlation measurements}

All $B$-mode diagnostics in this analysis use the discrete shear catalog $\mathcal{C} = \{(\gamma_i, w_i, \hat{\mathbf{n}}_i)\}_{i=1}^N$, comprising complex ellipticities $\gamma_i$, shear weights $w_i$ (Paper~\papercatalog), and sky positions $\hat{\mathbf{n}}_i$ for $N$ source galaxies.
The ellipticities $\gamma_i$ are shear estimates calibrated via metacalibration \citep{huff.mandelbaum17, sheldon.huff17}: galaxy images are artificially sheared to measure the $2 \times 2$ shear response matrix $\mathbf{R}$, yielding $\boldsymbol{\gamma}_i = \mathbf{R}^{-1}(\mathbf{e}_i - \mathbf{c}_i)$ from the observed ellipticity $\mathbf{e}_i$ and additive bias $\mathbf{c}_i$.
The two-point shear correlation functions $\xi_\pm(\theta)$ are formed from products of tangential and cross-component ellipticities for galaxy pairs at separation $\theta$.
We estimate $\xi_\pm$ by weighting each pair by metacalibration weights and accumulating contributions in angular bins:
\begin{equation}
{\xi}_\pm(\theta_k) = \frac{\sum_{i \neq j} w_i w_j \, \Pi(\theta_{ij}, \theta_k) \left[\gamma_{\mathrm{t},i} \gamma_{\mathrm{t},j} \pm \gamma_{\times,i} \gamma_{\times,j}\right]}{\sum_{i \neq j} w_i w_j \, \Pi(\theta_{ij}, \theta_k)}.
\label{eq:xi_estimator}
\end{equation}
Here $\gamma_{\mathrm{t}}$ and $\gamma_{\times}$ are the tangential and cross components of the ellipticity in the coordinate frame defined by the pair-separation vector, $\Pi(\theta_{ij}, \theta_k)$ is the binning kernel selecting pairs with separations in bin $k$, and $\theta_{ij} = \arccos(\hat{\mathbf{n}}_i \cdot \hat{\mathbf{n}}_j)$.
We evaluate \cref{eq:xi_estimator} using the tree-based correlation package \texttt{TreeCorr} \citep{jarvis04}.
We bin into 20 logarithmic bins over $\theta = 1$--$250$~arcmin, matching the configuration-space data-vector binning adopted in Paper~\paperconfig{}.

In harmonic space, the analogous statistics are the angular power spectra $C_\ell^{EE}$ and $C_\ell^{BB}$.
We estimate these using the MASTER algorithm \citep{hivon.etal02}, which recovers the true $C_\ell$ by deconvolving the mode-coupling matrix induced by the survey geometry, as implemented in \texttt{NaMaster} \citep{alonso.etal19}.
Traditional implementations operate on pixelized maps, but \citet{wolz.etal25} introduced a catalog-based estimator that evaluates spherical harmonic transforms directly at source positions, avoiding pixelization artifacts.
The same weighted ellipticities from \cref{eq:xi_estimator} enter this estimator:
\begin{equation}
\tilde{C}_\ell^{\alpha\beta} = \sum_{i,j} w_i w_j \, \gamma_i^a \, \gamma_j^b \; \frac{1}{2\ell+1} \sum_m {}_2 Y_{\ell m}^{a\alpha *}(\hat{\mathbf{n}}_i) \; {}_2 Y_{\ell m}^{b\beta}(\hat{\mathbf{n}}_j),
\label{eq:cl_definition}
\end{equation}
where $a, b \in \{1, 2\}$ label the two ellipticity components, ${}_2 Y_{\ell m}^{a\alpha}$ are spin-2 spherical harmonics, and $\alpha, \beta \in \{E, B\}$.
We then decouple the measured bandpowers and correct for noise bias following \citet{wolz.etal25}, yielding unbiased harmonic-space bandpower estimates $\hat{C}_\ell^{\alpha\beta}$.

\subsection{Pure $E$/$B$ decomposition}
\label{sec:pure_eb}

We decompose the measured correlation functions $\xi_\pm(\theta)$ into pure $E$-mode, $B$-mode, and ambiguous components following \citet{schneider.etal22}:
\begin{align}
\xi_+(\theta) &= \xi_+^{\mathrm{E}}(\theta) + \xi_+^{\mathrm{B}}(\theta) + \xi_+^{\mathrm{amb}}(\theta); \label{eq:decomp_plus} \\
\xi_-(\theta) &= \xi_-^{\mathrm{E}}(\theta) - \xi_-^{\mathrm{B}}(\theta) + \xi_-^{\mathrm{amb}}(\theta). \label{eq:decomp_minus}
\end{align}
Ambiguous modes $\xi_\pm^{\mathrm{amb}}(\theta)$ arise from shear patterns that cannot be assigned unique $E$- or $B$-mode origin on a finite angular interval; they take the functional forms $\xi_+^{\mathrm{amb}}(\theta) = a + b\theta^2$ and $\xi_-^{\mathrm{amb}}(\theta) = c\theta^{-2} + d\theta^{-4}$.
They therefore appear on large scales in $\xi_+$ and on small scales in $\xi_-$, as seen in \cref{fig:pure_eb_decomposition}.
The pure-mode correlation functions are computed from integral transforms over the measured $\xi_\pm(\theta)$:
\begin{align}
\xi_+^{\mathrm{E,B}}(\theta) &= \frac{1}{2}\left[\xi_+(\theta) \pm \xi_-(\theta) \pm \int_\theta^{\theta_{\max}} \frac{\mathrm{d}\theta'}{\theta'}\,
\xi_-(\theta')\left(4-\frac{12\theta^2}{\theta'^2}\right)\right] \nonumber \\
&\quad - \frac{1}{2}\left[S_+(\theta) \pm S_-(\theta)\right]; \label{eq:xip_EB} \\
\xi_-^{\mathrm{E,B}}(\theta) &= \frac{1}{2}\left[\xi_+(\theta) \pm \xi_-(\theta) + \int_{\theta_{\min}}^\theta \frac{\mathrm{d}\theta'\,\theta'}{\theta^2}\,
\xi_+(\theta')\left(4-\frac{12\theta'^2}{\theta^2}\right)\right] \nonumber \\
&\quad - \frac{1}{2}\left[V_+(\theta) \pm V_-(\theta)\right]. \label{eq:xim_EB}
\end{align}
Here, the $+$ sign gives $E$-modes and the $-$ sign gives $B$-modes.
The auxiliary functions $S_\pm(\theta)$ and $V_\pm(\theta)$ enforce boundary conditions that remove ambiguous modes: $\xi_+^{\mathrm{amb}} = S_+$ and $\xi_-^{\mathrm{amb}} = V_-$ \citep{schneider.etal22}.

The integral transforms require finely binned $\xi_\pm$ over an extended angular range.
We evaluate these transforms on a separate 1000-bin logarithmic grid spanning $\theta = 0.5$--$300$~arcmin; the extended range reduces edge effects in the boundary functions $S_\pm$ and $V_\pm$.
We apply the \citet{schneider.etal22} filter kernels to this grid using the \texttt{cosmo\_numba}\footnote{\url{https://github.com/aguinot/cosmo-numba}} implementation \citep{guinot.etal26}.
\Cref{fig:pure_eb_decomposition} shows the pure $E$/$B$ decomposition for the fiducial catalog.

\subsection{COSEBIs}
\label{sec:cosebis}

\begin{figure}[!tb]
\centering
\includegraphics[width=\columnwidth]{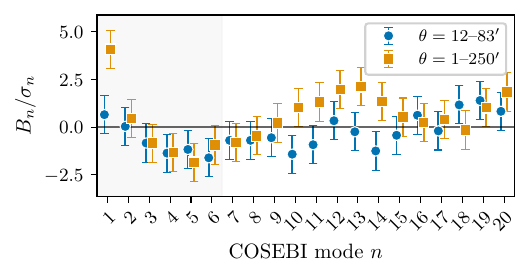}
\caption{COSEBI $B$-mode amplitudes for the fiducial catalog, normalized by uncertainty ($B_n/\sigma_n$).
Orange squares: full angular range $1$--$250$~arcmin; blue circles: fiducial scale cuts $\ebthetaXipMin$--$\ebthetaXipMax$~arcmin.
The shaded region highlights mode numbers $n \leq 6$, which capture nearly all cosmological information.
On the full range, $n=1$ shows a $>4\,\sigma$ excess, with oscillatory structure through all twenty modes, consistent with repeating additive shear bias (\cref{sec:discussion}).
Scale cuts suppress the first-mode excess to below \num{1}\,$\sigma$ and largely eliminate the oscillatory signature (see \cref{tab:pte_results}).}
\label{fig:cosebis_fiducial}
\end{figure}

COSEBIs compress $\xi_\pm(\theta)$ into discrete orthogonal modes $E_n$ and $B_n$, with clean $E$/$B$ separation on finite angular intervals \citep{schneider.eifler.krause10}.
The filter functions $T_{\pm n}(\theta)$ are constructed to satisfy boundary conditions that remove the ambiguous modes described in \cref{sec:pure_eb}.
Because the $T_{\pm n}$ are orthonormal, different modes capture largely independent information.
Each $T_{\pm n}(\theta)$ oscillates $n+1$ times across the angular interval; higher-$n$ modes are therefore sensitive to increasingly oscillatory features in $\theta$-space.
Conversely, a feature localized in angle projects onto many modes at once, producing an oscillatory signature across the basis, as in the full-range measurement of \cref{fig:cosebis_fiducial}.

We compute the COSEBI mode amplitudes $X_n \in \{E_n, B_n\}$ over a given angular range $[\theta_{\mathrm{min}}, \theta_{\mathrm{max}}]$~arcmin using logarithmic filter functions:
\begin{equation}
X_n = \frac{1}{2} \int_{\theta_{\mathrm{min}}}^{\theta_{\mathrm{max}}} \mathrm{d}\theta \, \theta \left[ T_{+n}(\theta) \xi_+(\theta) \pm T_{-n}(\theta) \xi_-(\theta) \right],
\label{eq:cosebi_xn}
\end{equation}
where the upper (lower) sign gives $E_n$ ($B_n$).
We evaluate these integrals on the same 1000-bin grid used for the pure-mode transforms (\cref{sec:pure_eb}).
As a convergence check, we repeated the calculation on a 10\,000-bin grid for the fiducial catalog and found the $B_n$ amplitudes and null-test PTEs to be stable.

\Cref{fig:cosebis_fiducial} shows COSEBI $B$-mode amplitudes for the first $n_{\mathrm{max}} = 20$ modes of the fiducial catalog.
For the UNIONS data vector, the $E$-mode signal saturates by $n \approx 5$--$6$ (see \cref{sec:discussion}), so we use $n \leq 6$ as the set of modes carrying most of the cosmological information.
We also report $B$-mode probability-to-exceed (PTE) statistics for $n \leq 20$; at full angular range, the oscillatory pattern extends across the full mode range, characteristic of the repeating additive shear bias identified by \citet{asgari.etal19a} (see \cref{sec:discussion}).

The COSEBI coefficients can reconstruct pure-mode correlation functions (\cref{sec:pure_eb}):
\begin{equation}
\xi_\pm^{X}(\theta) = \frac{\bar{\theta}^2}{\beta_\theta} \sum_{n=1}^{n_{\mathrm{max}}} X_n \, T_{\pm n}(\theta),
\label{eq:xi_XEB_reconstruction}
\end{equation}
where $\bar{\theta} = (\theta_{\mathrm{min}} + \theta_{\mathrm{max}})/2$ and $\beta_\theta = (\theta_{\mathrm{max}} - \theta_{\mathrm{min}})/(\theta_{\mathrm{max}} + \theta_{\mathrm{min}})$.
In practice, we compute the pure-mode correlation functions directly from the finely binned $\xi_\pm$ measurements rather than reconstructing them from the COSEBI coefficients.
COSEBIs require consistent angular ranges for $\xi_+$ and $\xi_-$, while the pure-mode integral transforms \cref{eq:xip_EB,eq:xim_EB} permit independent scale cuts for each.

The COSEBI modes can equivalently be computed from harmonic-space power spectra:
\begin{equation}
X_n = \sum_i \frac{\Delta\ell_i \, \ell_i}{2\pi} \, W_n(\ell_i) \, C_{\ell_i}^{XX},
\label{eq:cosebi_harmonic}
\end{equation}
where $W_n(\ell)$ are the harmonic-space COSEBI filter functions (Fourier duals of $T_{\pm n}(\theta)$, computed via FFT-log) and the sum runs over 96 square-root-spaced bandpower bins.
Each $W_n(\ell)$ oscillates $n+1$ times, mirroring its real-space counterpart.
The number of bandpower bins balances two requirements: too few cannot resolve the filter oscillations at high $n$, while too many lead to an unstable mode-coupling matrix deconvolution.
We chose 96 bins and validated the transform on GLASS simulations, confirming that modes $n \leq 6$---the range carrying nearly all cosmological $E$-mode information---agree between the two paths.
The comparison on data is presented in \cref{sec:discussion} and \cref{fig:harmonic_config_cosebis_full}.

\subsection{Catalog-based harmonic-space power spectra}
\label{sec:catalog_cls}

\begin{figure}[tb]
\centering
\includegraphics[width=\columnwidth]{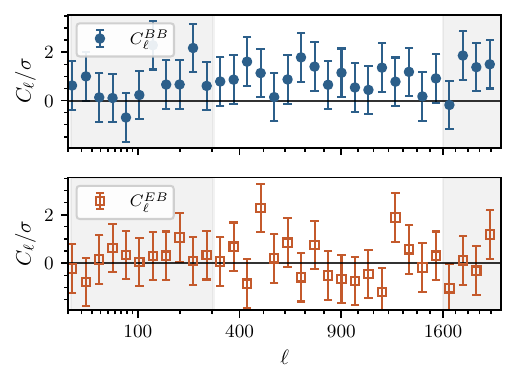}
\caption{Harmonic-space power spectra for the fiducial catalog, normalized by uncertainty.
Top: $B$-mode auto-power $C_\ell^{BB}/\sigma$. Bottom: $E$-$B$ cross-power $C_\ell^{EB}/\sigma$.
$C_\ell^{BB}$ is predominantly positive across the fiducial range.
$C_\ell^{EB}$ scatters symmetrically around zero, suggesting that any $B$-mode contamination is not strongly correlated with the lensing signal.}
\label{fig:cl_fiducial}
\end{figure}

We estimate $C_\ell^{EB}$ and $C_\ell^{BB}$ \cref{eq:cl_definition} directly from the discrete catalog, providing $\ell$-dependent $B$-mode diagnostics that complement the angular-space statistics.
The \citet{wolz.etal25} estimator represents the survey mask as delta functions at source positions and computes the spin-2 transforms on this irregular grid using the \texttt{ducc} library \citep{reinecke.seljebotn23}.
We subtract an analytic shot-noise contribution following the implementation in Paper~\paperharmonic.
We bin into 32 bandpowers with square-root spacing ($\ell^{0.5}$) over $8 \leq \ell \leq 2048$.
\Cref{fig:cl_fiducial} shows the measured $C_\ell^{BB}$ and $C_\ell^{EB}$ for the fiducial catalog.

\subsection{Covariance estimation}
\label{sec:covariance}

For each $B$-mode statistic, we start from the same covariance prescriptions adopted in the companion cosmology papers.
In configuration space, Paper~\paperconfig{} uses \texttt{CosmoCov} \citep{krause.eifler17, fang.eifler.krause20} to model the covariance of the shear correlation functions $\xi_\pm$.
In harmonic space, Paper~\paperharmonic{} uses the Gaussian covariance from \texttt{NaMaster}'s iNKA framework for the $EB$ and $BB$ spectra, with separate non-Gaussian terms added only for the $EE$ cosmology analysis.
Here we use those same inputs to construct the covariances needed for the derived $B$-mode data vectors.

For $C_\ell^{BB}$ and $C_\ell^{EB}$, the harmonic-space covariance is used directly.
Paper~\paperharmonic{} validates this $B$-mode covariance against 10\,000 Gaussian simulations and finds good agreement away from the lowest multipoles (Appendix~C therein).
For the configuration-space statistics, we begin from the finely binned \texttt{CosmoCov} covariance of $\xi_\pm(\theta)$ on a 1000-bin logarithmic grid spanning $\theta = 0.5$--$300$~arcmin, using the survey properties of Paper~\papercatalog{} and the Planck 2018 fiducial cosmology \citep{planck20}.

For COSEBIs, the covariance follows directly from the linear transformation
\begin{equation}
\mathbf{C}_{B_n B_m} = \mathbf{T}^{\mathsf{T}} \mathbf{C}_{\xi} \mathbf{T},
\end{equation}
where $\mathbf{T}$ encodes the COSEBI filter functions.
For the pure-mode correlation functions, no equivalent direct transformation is available.
Instead, we draw $N_{\mathrm{samples}} = 2000$ realizations of $\xi_\pm(\theta)$ from the same finely binned \texttt{CosmoCov} covariance, propagate each realization through the integral transforms in \cref{eq:xip_EB,eq:xim_EB}, and compute the empirical covariance of the reconstructed $\xi_\pm^{\mathrm{E/B}}$.
This captures the cross-covariance between $\xi_+$ and $\xi_-$, correlations between $E$/$B$/ambiguous mode types, and angular-bin correlations induced by the overlapping filter support.
When computing $\chi^2$ statistics for PTE assessments, we apply the correction factor $(N_{\mathrm{samples}} - N_{\mathrm{obs}} - 2)/(N_{\mathrm{samples}} - 1)$ to the inverse covariance of the Monte Carlo pure-mode estimate \citep{hartlap07};
for $N_{\mathrm{samples}} = 2000$ and the data-vector dimensions used here ($N_{\mathrm{obs}} \leq 40$), this amounts to at most a 2\% correction.

\begin{figure}[tb]
\centering
\includegraphics[width=\columnwidth]{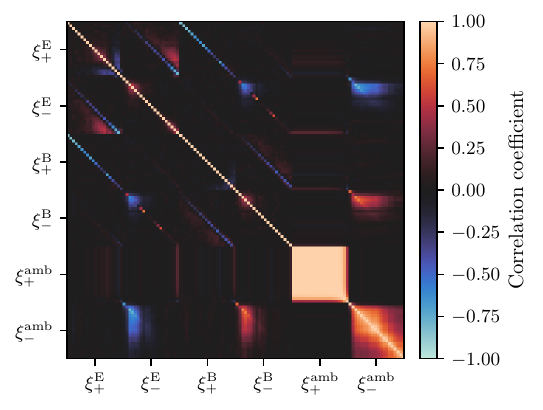}
\caption{Pure $E$/$B$-mode correlation-function covariance matrix for the fiducial catalog from semi-analytical propagation of 2000 Monte Carlo realizations.
The matrix is organized into six blocks: $\xi_+^{\mathrm{E}}$, $\xi_-^{\mathrm{E}}$, $\xi_+^{\mathrm{B}}$, $\xi_-^{\mathrm{B}}$, $\xi_+^{\mathrm{amb}}$, $\xi_-^{\mathrm{amb}}$, with 20 angular bins per block.
Off-diagonal correlations arise from integral coupling in the pure-mode filters.}
\label{fig:eb_covariance}
\end{figure}

For all $B$-mode null tests, we retain only the Gaussian part of the covariance.
In configuration space, this keeps the calculation tractable on the finely binned integration grid, and in harmonic space it matches the Gaussian iNKA covariance used for the $BB$ and $EB$ null tests in Paper~\paperharmonic.
This approximation underestimates the true variances because it omits connected trispectrum and super-sample covariance terms.
The reported PTEs are therefore conservative: a null test that passes with Gaussian-only covariance would generally pass with a more complete covariance estimate.
\Cref{fig:eb_covariance} shows the pure $E$/$B$ covariance structure.

\begin{table*}[!bt]
  \centering
  \caption{$B$-mode PTE values across catalog versions at fiducial and full-range scale cuts.
  Fiducial scale cuts are $\ebthetaXipMin$--$\ebthetaXipMax$~arcmin in configuration space and $300 < \ell < 1600$ in harmonic space; the full range is $1$--$250$~arcmin / $8 \leq \ell \leq 2048$.
  Only the fiducial catalog passes all statistics at the adopted cuts.
  Bold values indicate PTE $< 0.05$ (null-test failure).}
  \label{tab:pte_results}
\begin{tabular}{ll cc @{\hskip 8pt} ccc @{\hskip 8pt} c}
    \hline\hline
    \noalign{\vskip 3pt}
    & & \multicolumn{2}{c}{COSEBIs} & \multicolumn{3}{c}{Pure E/B} & $C_\ell$ \\
    \cmidrule(lr){3-4} \cmidrule(lr){5-7} \cmidrule(l){8-8}
    Version & Scale cuts & $B_n$ ($n \leq 6$) & $B_n$ ($n \leq 20$) & $\xi_+^{\mathrm{B}}$ & $\xi_-^{\mathrm{B}}$ & $\xi_{\mathrm{tot}}^{\mathrm{B}}$ & $C_\ell^{BB}$ \\
    \hline
    \noalign{\vskip 3pt}
    Initial & Fiducial & \num{0.94} & \num{0.61} & \num{0.51} & \textbf{\num{0.005}} & \textbf{\num{0.030}} & \textbf{\num{0.003}}  \\
     & Full range & \textbf{\num{7.37e-10}} & \textbf{\num{4.15e-07}} & \textbf{\num{0.026}} & \textbf{\num{7.71e-04}} & \textbf{\num{0.004}} & \textbf{\num{2.53e-04}}  \\
    \noalign{\vskip 2pt}
    Size cuts (fiducial) & Fiducial & \num{0.78} & \num{0.94} & \num{0.31} & \num{0.26} & \num{0.18} & \num{0.30}  \\
     & Full range & \textbf{\num{1.37e-05}} & \textbf{\num{1.08e-04}} & \num{0.45} & \num{0.16} & \num{0.40} & \num{0.13}  \\
    \noalign{\vskip 2pt}
    Masked & Fiducial & \num{0.60} & \num{0.81} & \textbf{\num{0.020}} & \textbf{\num{0.047}} & \textbf{\num{0.014}} & \num{0.43}  \\
     & Full range & \textbf{\num{4.94e-06}} & \textbf{\num{2.50e-05}} & \textbf{\num{0.039}} & \textbf{\num{0.020}} & \textbf{\num{0.018}} & \num{0.11}  \\
    \noalign{\vskip 2pt}
    Relaxed flags & Fiducial & \num{0.82} & \num{0.83} & \num{0.78} & \textbf{\num{9.40e-04}} & \textbf{\num{0.014}} & \textbf{\num{5.74e-10}}  \\
     & Full range & \textbf{\num{1.70e-10}} & \textbf{\num{1.16e-08}} & \num{0.10} & \textbf{\num{2.94e-06}} & \textbf{\num{9.07e-04}} & \textbf{\num{1.78e-12}}  \\
  \hline
\end{tabular}
\end{table*}

\subsection{$B$-mode significance}
\label{sec:scale_cuts}

We assess $B$-mode significance across a grid of angular scale-cut combinations.
For each scale range, we compute a $\chi^2$ statistic testing the $B$-mode data vector $\vec{d}_B$ against zero:
\begin{equation}
\chi^2_B = \vec{d}_B^{\mathsf{T}} \, \mathbf{C}_{BB}^{-1} \, \mathbf{d}_B,
\label{eq:chi2_bmode}
\end{equation}
where $\mathbf{C}_{BB}$ is the $B$-mode covariance submatrix for the selected scale range (corrected for finite-sample bias in the MC-propagated pure-mode covariance; \cref{sec:covariance}).
The probability-to-exceed (PTE) is then $P(\chi^2 > \chi^2_B \,|\, \nu)$ for $\nu$ degrees of freedom equal to the number of data points in the range.
Since the scale-cut boundaries are themselves selected using the PTE grid, one could treat them as two fitted parameters and reduce $\nu$ accordingly.\footnote{Doing so lowers the minimum PTE across all statistics from 0.18 to 0.09, still above the 0.05 threshold.}
For the pure-mode decomposition, we compute PTEs separately for $\xi_+^{\mathrm{B}}$ and $\xi_-^{\mathrm{B}}$, as well as a joint test $\xi_{\mathrm{tot}}^{\mathrm{B}}$ using the concatenated data vector and full cross-covariance.
For COSEBIs, we evaluate PTEs using both $n \leq 6$ and $n \leq 20$ (\cref{sec:cosebis}).
For $C_\ell^{BB}$, we apply the same framework across multipole ranges.
We adopt a uniform threshold of PTE $=0.05$ for all tests.
We do not attempt an explicit correction for look-elsewhere effects across the many scale-cut combinations, because the tests are strongly correlated both within and across statistics; we instead require the adopted cuts to lie in broad stable regions that pass across all three frameworks.

\subsection{Validation on simulations}
\label{sec:mock_validation}

We validate the analysis pipeline on 350 lognormal simulated catalogs generated with the \texttt{GLASS} package \citep{tessore.etal23}, generated with the same survey footprint, galaxy positions, weights, and Planck~2018 cosmology as the UNIONS data.
These simulated catalogs contain no PSF-related systematics, so they provide a direct check that our estimators do not generate spurious $B$ modes when applied to a noise-dominated signal.
Across the ensemble, the pure-mode decomposition, COSEBIs, and harmonic-space power spectra all recover $B$ modes consistent with zero;
for COSEBIs specifically, the mean $B_n$ over a subset of 100 realizations is biased by less than $0.3\,\sigma$ per mode.
We also find that the scatter in the simulated realizations is comparable to the fiducial error bars adopted for each statistic, indicating that our uncertainty estimates are of the right order.

The companion cosmology papers validate their pipelines with the same simulation suite: Paper~\paperconfig{} checks the configuration-space pipeline and \texttt{CosmoCov} covariance, while Paper~\paperharmonic{} checks the harmonic-space pipeline and \texttt{NaMaster} covariance.
Here we apply the simulations to the $B$-mode statistics specifically, complementing the cosmological-inference tests in those papers.
Taken together, these tests show that the low-level $B$-mode structure discussed below is a property of the data rather than an artifact of the estimators or their numerical implementation.

\section{Results}
\label{sec:results}

\begin{figure*}[p]
\centering
\includegraphics[width=0.88\textwidth]{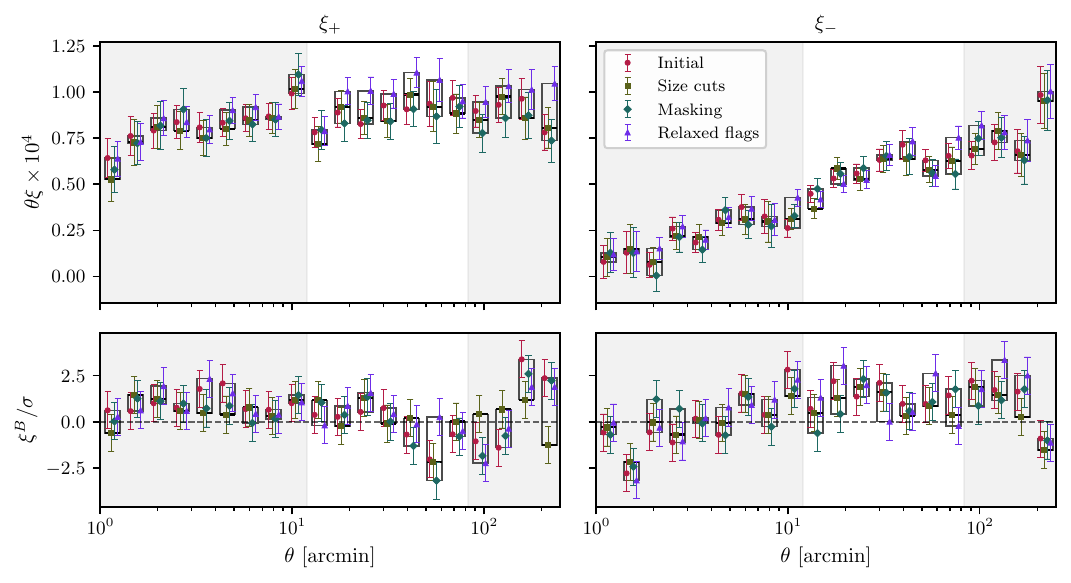}\\[1pt]
\includegraphics[width=0.88\textwidth]{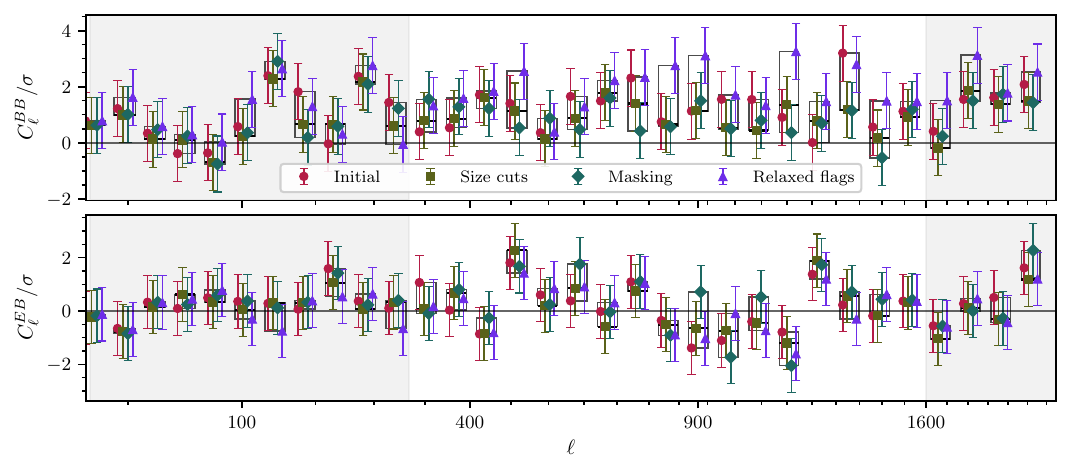}\\[1pt]
\includegraphics[width=0.88\textwidth]{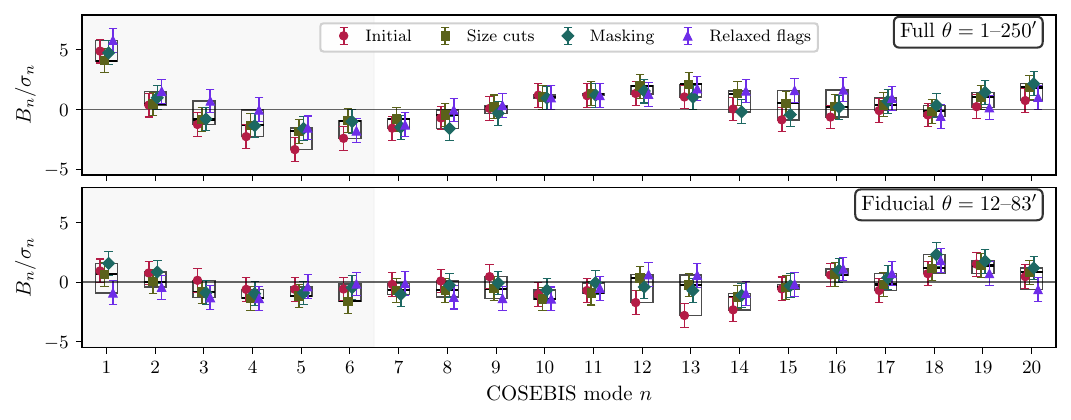}
\caption{$B$-mode data vectors across catalog versions: initial (red circles), size-cut/fiducial (gold squares), masked (teal diamonds), and relaxed-flags (purple triangles).
Rectangles delineate each bin in the horizontal direction and span the range of values across versions in the vertical direction; horizontal lines mark the fiducial value.
Shaded bands mark fiducial scale cuts for $\xi_\pm$ and $C_\ell$.
\textit{Top:} Pure $E$/$B$-mode correlation functions.
Upper panels show $\xi_+$ and $\xi_-$ scaled by $\theta$; lower panels show $B$-mode components $\xi_\pm^{\mathrm{B}}/\sigma$.
\textit{Middle:} Harmonic-space power spectra $C_\ell^{BB}/\sigma$ and $C_\ell^{EB}/\sigma$.
\textit{Bottom:} COSEBI $B$-mode amplitudes $B_n/\sigma_n$ at full range (upper) and fiducial cuts (lower).
Only the size-cut catalog passes all statistics at the fiducial scale cuts (\cref{tab:pte_results}).}
\label{fig:pure_eb_versions}
\label{fig:cl_versions}
\label{fig:cosebis_versions}
\end{figure*}

\begin{figure*}[p]
\centering
\includegraphics[width=\textwidth]{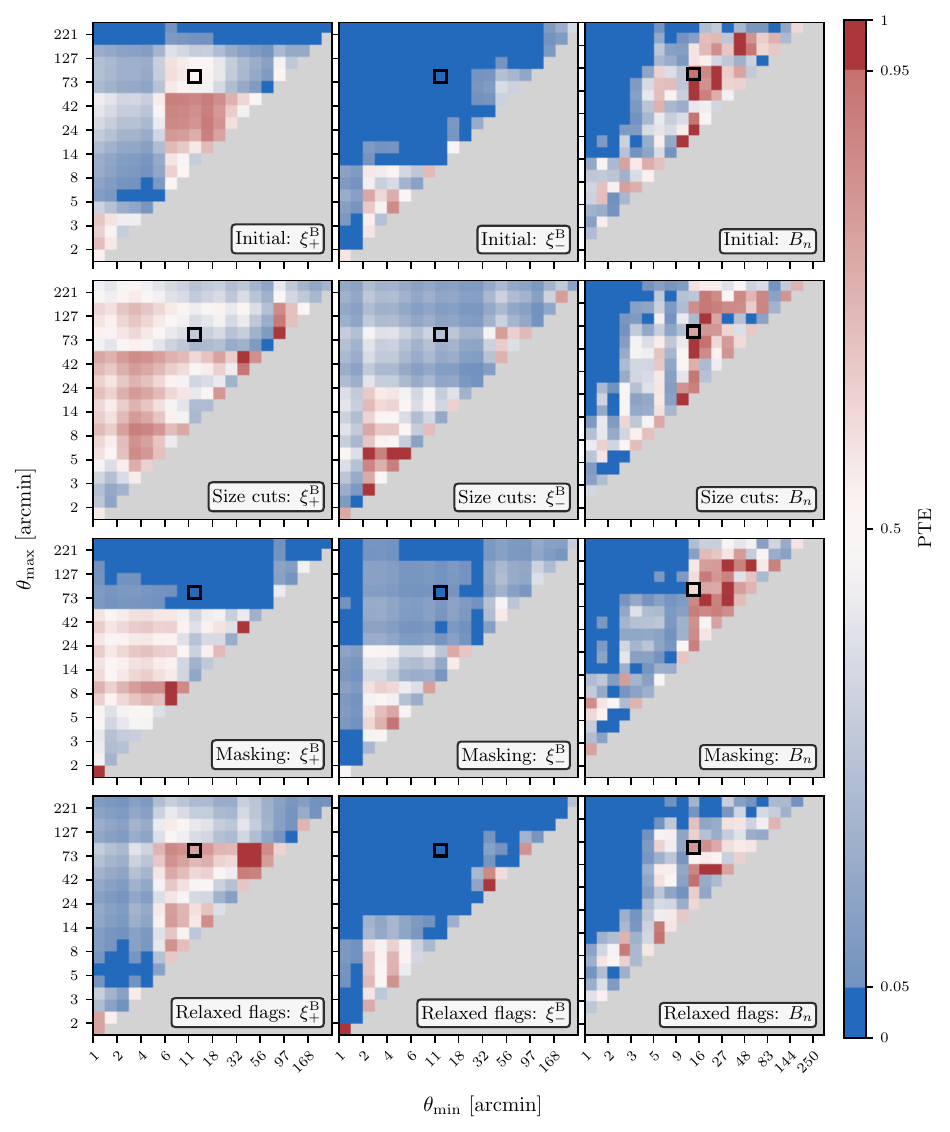}
\caption{Configuration-space PTE maps for all four catalog versions (Paper~\papercatalog, Table~H.1).
Columns show $\xi_+^{\mathrm{B}}$, $\xi_-^{\mathrm{B}}$, and COSEBI $B_n$ PTEs as a function of angular scale cuts; rows show different catalog versions.
Solid blue cells indicate PTE~$< 0.05$; whites and reds indicate consistency with zero.
Black squares mark the adopted cuts ($\ebthetaXipMin$--$\ebthetaXipMax$~arcmin).
The size cut (fiducial versus initial) expands the acceptance region, while stellar masking contracts it.}
\label{fig:pte_heatmaps}
\end{figure*}

\begin{figure*}[!t]
\centering
\includegraphics[width=0.95\textwidth]{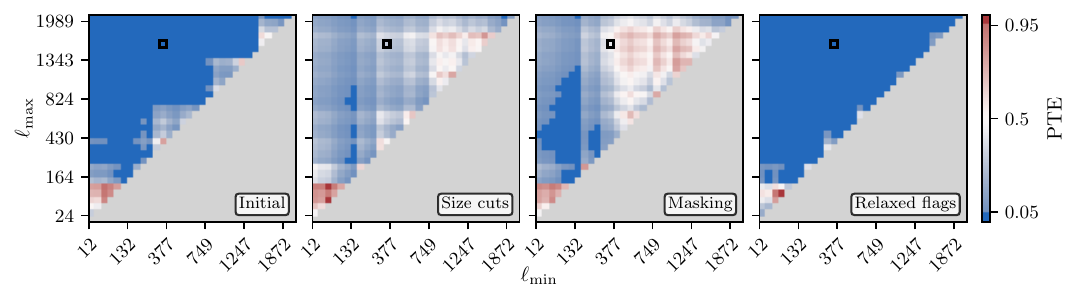}
\caption{Harmonic-space $C_\ell^{BB}$ PTE maps for all four catalog versions (Paper~\papercatalog, Table~H.1).
Solid blue cells indicate failing PTEs; whites and reds indicate consistency with zero.
The initial and relaxed-flags catalogs show widespread failures across multipole ranges;
the fiducial and masked catalogs pass across nearly all multipole combinations.}
\label{fig:pte_cl}
\end{figure*}

For the fiducial catalog before scale cuts, COSEBIs fail with a $>4\,\sigma$ first mode, while the pure $B$-mode correlation functions and $C_\ell^{BB}$ pass but show low-level $B$-mode structure (\cref{tab:pte_results}).
At the adopted scale cuts, all null tests pass.
We chose those cuts using the $B$-mode tests together with additional systematics checks and blinded inference-stability checks in Papers~\paperconfig{} and~\paperharmonic; they are more conservative than the PTEs alone would require.
The object-wise PSF-leakage correction (Paper~\papercatalog) shifts individual PTEs by $\lesssim 0.05$ across all catalog versions and statistics; no pass/fail conclusion changes.
We report leakage-corrected results throughout.

The low-level features that drive the full-range behavior leave different signatures in each statistic.
In the pure-mode decomposition (\cref{fig:pure_eb_decomposition}), $\xi_+^{\mathrm{B}}/\sigma$ shows a broad positive excess on the smallest angular scales that tapers toward zero by $\theta \sim \SI{10}{arcmin}$;
the corresponding $\xi_-^{\mathrm{B}}$ is quieter on the smallest scales but remains broadly elevated from roughly \SI{7}{arcmin} to the largest scales.
The COSEBI data vector (\cref{fig:cosebis_fiducial}) compresses the same structure into a smaller number of modes: on the full angular range, the first mode exceeds $4\,\sigma$ and the higher modes show coherent oscillations consistent with the repeating additive pattern discussed in \cref{sec:discussion};
after applying the fiducial angular cuts, the first-mode excess drops below \num{1}\,$\sigma$ and the oscillatory pattern largely disappears.
In harmonic space (\cref{fig:cl_fiducial}), $C_\ell^{BB}$ shows a low-level positive offset across much of the fiducial range, with the most conspicuous outliers near $\ell \approx 125$ and $250$, and above $\ell \approx 1600$.
Even so, the fiducial range $300 < \ell < 1600$ passes the null test, while $C_\ell^{EB}$ remains consistent with zero, suggesting that any $B$-mode contamination is not strongly correlated with the lensing signal.

\Cref{fig:pure_eb_versions} compares the $B$-mode measurements across catalog versions, and \cref{tab:pte_results} summarizes the PTEs.
Each statistic shows a different pattern of failures.
In configuration space, the initial and relaxed-flags catalogs fail in $\xi_-^{\mathrm{B}}$ and in $\xi_{\mathrm{tot}}^{\mathrm{B}}$ at the fiducial cuts.
The relaxed-flags catalog has a similar acceptance region to the fiducial in $\xi_+^{\mathrm{B}}$, but a narrower one in $\xi_-^{\mathrm{B}}$, with the failure boundary approaching the adopted cuts (PTE $= \configPteElevenThreeXim$).
The masked catalog fails all three pure-mode tests despite passing in harmonic space and COSEBIs ($n \leq 6$).
The PTE maps (\cref{fig:pte_heatmaps}) show that masking increases the $B$-mode significance on both the largest and smallest scales, although the $\xi_-^{\mathrm{B}}$ PTE falls only marginally below threshold ($0.047$).
In harmonic space (\cref{fig:pte_cl}), the initial and relaxed-flags catalogs fail across most multipole combinations, whereas the fiducial and masked catalogs show broad acceptance regions.

COSEBI $B_n$ tests with $n \leq 6$ pass for all four versions at fiducial cuts, despite pure-mode and harmonic-space failures in three of them (\cref{tab:pte_results}).
Only the fiducial catalog passes all statistics at the adopted cuts.
At full range, all versions show low-order oscillatory structure consistent with the repeating additive pattern discussed below, driving COSEBI ($n \leq 6$) PTEs to $10^{-5}$ or below; at fiducial cuts, all versions pass, although the initial catalog retains a mild oscillatory pattern.

\Cref{fig:pte_cl,fig:pte_heatmaps} map null-test performance over the full grid of scale-cut combinations.
In each representation, the adopted cuts sit well inside broad acceptable regions, so moving them by several bins in either direction does not change the outcome.
For COSEBIs, whose data vector is indexed by mode number rather than angle, the PTE heatmap translates null-test performance back into angular space.

\section{Discussion}
\label{sec:discussion}

The pure-mode correlation functions, COSEBIs, and harmonic-space power spectra presented in this paper weight angular scales and treat ambiguous modes differently, so they do not respond identically to the same contamination.
When they disagree, we treat that as evidence of contamination in the sample even if each statistic is not equally sensitive to it, and require the adopted catalog and cuts to pass in all three frameworks.

At full range, all four catalogs show a COSEBI first-mode excess exceeding $4\,\sigma$ and oscillatory structure across all twenty modes.
Because COSEBIs are orthogonal over a finite angular range (\cref{sec:cosebis}), a feature compact in angle spreads across the mode spectrum; the full-range oscillations are qualitatively consistent with contamination at CCD angular scales (\cref{sec:results}, \cref{fig:cosebis_fiducial}).
In the pure-mode and harmonic-space data vectors, no comparably localized feature is apparent: $\xi_+^{\mathrm{B}}$ shows mild excess at the smallest and largest separations, and $C_\ell^{BB}$ has a few ${\gtrsim}2\,\sigma$ outliers at low $\ell$ and elevated power in the highest multipole bins, but neither displays a sharp angular signature that would point unambiguously to CCD-scale contamination.
MegaCam's 40 CCDs span $6$--$14$~arcmin individually, and our adopted lower cut of \ebthetaXipMin~arcmin excludes most pair separations at these scales; at the adopted scale cuts, the COSEBI features are suppressed and all four catalogs pass.
Pure-mode correlation functions and $C_\ell^{BB}$ on the other hand pass at the adopted cuts only for the size-cut catalog and for the size-cut and masked versions, respectively.
These equivocal results lead us to argue against relying on any single null test; taken together, we choose the size-cut catalog and our adopted scale cuts as the combination least affected by systematic contamination for cosmological inference in Papers~\paperconfig{} and~\paperharmonic.

\begin{figure*}[!t]
\centering
\includegraphics[width=\textwidth]{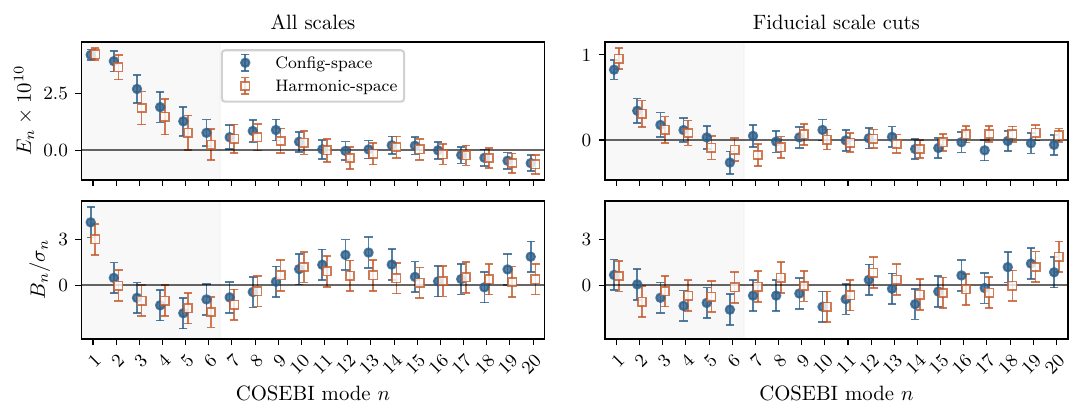}
\caption{COSEBI $E_n$ (top row) and $B_n/\sigma_n$ (bottom row) for the fiducial catalog, computed from configuration-space $\xi_\pm$ (filled circles) and harmonic-space bandpowers (open squares).
Gray shading marks modes $n \leq 6$, where the cosmological information is concentrated and we confirm consistency between the two paths on GLASS simulations.}
\label{fig:harmonic_config_cosebis_full}
\end{figure*}

Although the masked catalog passes in harmonic space and COSEBIs ($n \leq 6$), this ostensibly more conservative choice fails all three pure-mode tests.
Excising stellar halos introduces irregular gaps in the survey footprint, and because $B$-modes are non-local, a more complex mask geometry may increase the fraction of modes classified as ambiguous, leaving the remaining pure-mode estimates less stable; alternatively, removing regions of elevated shape noise near bright stars could increase the significance of contamination elsewhere.
We cannot distinguish these explanations with the present data.
The initial and relaxed-flags catalogs fail in $\xi_-^{\mathrm{B}}$ and harmonic space, consistent with poorly resolved or deblended objects contributing to the contamination; the relaxed-flags catalog also shifts the total $\xi_+$ systematically higher than the other versions beyond \SI{16}{arcmin} (\cref{fig:pure_eb_versions}), suggesting that admitting these objects affects the signal itself, not only the $B$-mode diagnostics.

Configuration-space and harmonic-space estimators respond differently to the same contamination.
Pure-mode correlation functions are localized in angular separation, so cuts that remove CCD-scale separations directly reduce the contamination.
In multipole space, a feature narrow in angle maps onto a broad range of $\ell$, so it cannot be isolated by cutting a localized multipole range.
COSEBIs apply filter functions to the two-point statistics that concentrate cosmological information into the first few modes; contamination at a characteristic angular scale spreads across the mode spectrum rather than isolating in a single mode.
Pure-mode and COSEBI filters also discard ambiguous modes (\cref{sec:pure_eb}) that cannot be uniquely classified as $E$ or $B$ on a finite angular interval.
The harmonic-space estimator used here does not deproject ambiguous modes; it assigns all power to $C_\ell^{EE}$ or $C_\ell^{BB}$, and any residual mask-induced $E$-to-$B$ leakage after mode-coupling deconvolution is absorbed into $C_\ell^{BB}$.

We show directly that $C_\ell^{BB}$ and COSEBIs have different sensitivity to systematic features, even when computed from the same data, by transforming the harmonic-space bandpowers into COSEBIs (\cref{eq:cosebi_harmonic}).
The $C_\ell^{BB}$ power spectrum passes the null test at our fiducial multipole range, yet COSEBIs computed from the same $C_\ell^{BB}$ fail at full angular range---because the COSEBI filter functions $W_n(\ell)$ weight the power spectrum differently, concentrating sensitivity on the scales where the contamination resides.
This provides empirical support for the argument of \citet{asgari.etal19a} that a null test in one representation cannot validate another; \citet{jefferson.etal25} reached the same conclusion across DES-Y3, KiDS-1000, and HSC-Y3, finding that tomographic bins passing pseudo-$C_\ell$ $B$-mode tests can fail when reanalyzed with \textsc{HybridEB}.
The converse also holds: a $B$-mode failure in any statistic should raise concern for cosmological inference in all of them, since the underlying contamination is present in the data regardless of how it is projected.

\Cref{fig:harmonic_config_cosebis_full} compares COSEBIs computed from configuration-space $\xi_\pm$ and from harmonic-space bandpowers.
Both $E_n$ and $B_n$ agree across the two computation paths for all twenty modes, with the tightest correspondence in the first six modes validated on GLASS simulations.
At fiducial scale cuts, both paths yield $B$-modes consistent with zero (PTE $= \harmCosebisPteSixThreeFid$ harmonic, $\cfgCosebisPteSixThreeFid$ configuration; $n \leq 6$).
At full range, both detect the same structure (PTE $= \harmCosebisPteSixThreeFull$ and $\cfgCosebisPteSixThreeFull$).
This confirms that the sensitivity to systematic contamination is set by the filter functions, not by the basis in which the two-point function is measured.
Throughout this analysis, we use Gaussian-only covariance, which underestimates variances for the UNIONS geometry (\cref{sec:covariance}); the reported PTEs are therefore conservative.

\citet{asgari.etal19a} identified an oscillatory COSEBI $B$-mode pattern in multiple Stage-III surveys---CFHTLenS, KiDS-450, and DES-SV---tracing it to a repeating additive shear bias (i.e.\ a spatially varying $c$-term) from detector-level effects fixed in focal-plane coordinates.
Its appearance across three Stage-III surveys points to a detector-level phenomenon rather than a pipeline-specific artifact, a point especially relevant for UNIONS because CFHTLenS also used MegaCam on CFHT \citep{heymans.etal12, guinot.etal22}.
The dither strategy, however, differs substantially.
CFHTLenS and KiDS used small dithers designed to bridge chip gaps, so the focal-plane pattern maps almost directly onto the sky.
UNIONS dithers by ${\sim}1/3$ of the MegaCam field of view \citep{gwyn.etal25}, larger than a single CCD, so each sky position averages the additive bias from several different CCDs across exposures.
\citet{asgari.etal19a} note that the $B$-mode signature of a repeating additive pattern depends on the dithering strategy, pointing to DES-SV's half-field dithers as one reason its pattern differs from the small-dither surveys.
The larger UNIONS dithers may similarly attenuate the amplitude of the repeating pattern, but are unlikely to fully eliminate it: the effective pattern is the per-CCD bias convolved with the dither geometry, potentially spreading the characteristic angular scale beyond the $6$--$14$~arcmin CCD footprint.

Several mechanisms can create additive shear structure at CCD angular scales.
The $\rho$-statistics and $\xi_{\mathrm{sys}}$ diagnostics in Paper~\papercatalog{} show scale-dependent PSF residuals with a feature near the ${\sim}10$~arcmin MegaCam CCD scale, consistent with per-CCD PSF contributions; astrometric residuals may also play a role.
Leakage correction has negligible impact on $B$-mode PTEs at the adopted scales (\cref{sec:results}).
However, the adopted cuts themselves exclude the large scales where PSF leakage is strongest, so this does not rule out leakage as a contributor to the full-range signal.
At the fiducial scales, residual contamination from higher-order PSF moments or other effects beyond a second-order leakage model may still contribute \citep{zhang.etal23}.

Paper~\papersims{} identifies a pipeline-specific source arising from the improper propagation of exposure offsets within the shape measurement pipeline.
In the current multi-exposure fitting procedure, objects are effectively assumed to be centered on the pixel grid of each individual exposure.
However, a consistent treatment would require recentering the model using the World Coordinate System (WCS), anchored to the detection position defined at the tile level.
In areas where CCDs overlap, the relative pixel offsets between exposures can induce a preferred orientation in the inferred galaxy shapes, corresponding to the vector displacements between exposure pixel grids.
As a result, a coherent additive shear pattern may be imprinted across the CCD overlap regions, on angular scales between the per-CCD footprint ($\sim 6$--$14$~arcmin) and the UNIONS dither spacing ($1/3$ MegaCam FoV, or $\sim 19$~arcmin).
The observed $B$-mode contamination concentrates at $\theta_\mathrm{min} \lesssim 12$~arcmin (\cref{fig:pte_heatmaps}), close to the per-CCD scale but below the dither offsets.
Whether this mechanism is responsible for the observed $B$ modes remains an open question.
Even so, a contamination localized near the CCD scale would be cut away in configuration space, while its Hankel transform to harmonic space would spread across multipoles, beyond any single $\ell$ cut.
If a comparable additive contribution were present in the $E$ modes, it would lift the harmonic-space $S_8$ relative to configuration space, and could contribute to the mild offset seen between Papers~\paperconfig{} and~\paperharmonic.
Including configuration-space scales down to $5$~arcmin does move its $S_8$ toward the harmonic value (Paper~\paperconfig).
While a fraction of these effects may be absorbed by standard multiplicative bias corrections, we mitigate any remaining impact on cosmological constraints by cutting the majority of CCD scales; the adopted cuts pass without fine-tuning.

An alternative to excluding contaminated scales would be to model the $B$-mode signal as a bias and subtract it from the total measurement.
However, the relationship between $E$- and $B$-mode contamination depends on the physical mechanism; \citet{asgari.etal19a} showed that different systematics produce markedly different $E$/$B$ ratios, so the $B$-mode amplitude alone does not determine the $E$-mode contamination.
We therefore treat $B$-modes as a diagnostic rather than a correction: scale cuts that remove $B$-mode contamination also remove any corresponding contamination of the total signal, without assuming a model for the relationship between them.

\section{Conclusions}

UNIONS-3500 weak-lensing $B$ modes are consistent with zero at the adopted scale cuts across pure-mode correlation functions, COSEBIs, and harmonic-space power spectra.
The cuts lie in broad stable regions of the PTE maps.

On the full angular range, all catalog versions show an oscillatory COSEBI $B$-mode pattern consistent with repeating additive shear bias at CCD angular scales \citep{asgari.etal19a}. This is especially relevant for UNIONS because a similar signature also appeared in CFHTLenS, which likewise used MegaCam on CFHT \citep{heymans.etal12, guinot.etal22}.
Our adopted lower cut suppresses this pattern without discarding the angular scales that carry most of the cosmological signal.
By computing COSEBIs from the harmonic-space bandpowers, we confirm that the disagreement is not a matter of harmonic versus real space: $C_\ell^{BB}$ passes where COSEBIs fail, because the COSEBI filter functions concentrate sensitivity on the contaminated scales.

Of the four catalog variants tested, only the fiducial passes in every representation.
Each statistic responds differently: tightening the galaxy-size cut suppresses $B$-mode power in $\xi_-^{\mathrm{B}}$ and harmonic space; adding stellar-halo masks introduces new pure-mode failures; and COSEBIs ($n \leq 6$) pass for all versions, even though the full-range measurements show strong low-order structure.
These differences arise because the same contamination projects differently into angular, modal, and multipole representations; demanding that all three frameworks pass simultaneously places tighter constraints than any single null test.
Papers~\paperconfig{} and~\paperharmonic{} adopt these fiducial cuts for cosmological inference.

Because the underlying contamination is present in the data regardless of how it is projected, a $B$-mode failure in any one framework warrants scrutiny across all of them.
Rather than modeling $B$-modes as a bias to subtract, we use them as a diagnostic: scale cuts that remove $B$-mode contamination also remove any corresponding contamination of the cosmological signal, without assuming a model for the $E$/$B$ ratio.
Whether the same CCD-scale effects arise in \textit{Euclid} and the Legacy Survey of Space and Time (LSST), with their different detector geometries and dithering strategies, remains to be seen.
As surveys become systematics-limited, disagreements between $B$-mode estimators will carry as much information as their agreements; multi-framework validation will be needed to distinguish instrumental contamination, pipeline effects, and residual astrophysical signals.

\section*{Data availability}
A subset of the raw data underlying this article is publicly available via the Canadian Astronomical Data Centre at \url{http://www.cadc-ccda.hia-iha.nrc-cnrc.gc.ca/en/megapipe/}.
The remaining raw data and all processed data are available to members of the Canadian and French communities via reasonable requests to the principal investigators of the Canada-France Imaging Survey, Alan McConnachie and Jean-Charles Cuillandre.

\begin{acknowledgements}
We are honoured and grateful for the opportunity of observing the Universe from Maunakea and Haleakala, which both have cultural, historical and natural significance in Hawaii. This work is based on data obtained as part of the Canada-France Imaging Survey, a CFHT large program of the National Research Council of Canada and the French Centre National de la Recherche Scientifique. Based on observations obtained with MegaPrime/MegaCam, a joint project of CFHT and CEA Saclay, at the Canada-France-Hawaii Telescope (CFHT) which is operated by the National Research Council (NRC) of Canada, the Institut National des Science de l'Univers (INSU) of the Centre National de la Recherche Scientifique (CNRS) of France, and the University of Hawaii. This research used the facilities of the Canadian Astronomy Data Centre operated by the National Research Council of Canada with the support of the Canadian Space Agency. This research is based in part on data collected at Subaru Telescope, which is operated by the National Astronomical Observatory of Japan.
Pan-STARRS is a project of the Institute for Astronomy of the University of Hawaii, and is supported by the NASA SSO Near Earth Observation Program under grants 80NSSC18K0971, NNX14AM74G, NNX12AR65G, NNX13AQ47G, NNX08AR22G, 80NSSC21K1572 and by the State of Hawaii.
CD and MK acknowledge support from the Agence Nationale de la Recherche (ANR-22-CE31-0014-01) TOSCA project.
This work was made possible by utilizing the CANDIDE cluster at the Institut d'Astrophysique de Paris. The cluster was funded through grants from the PNCG, CNES, DIM-ACAV, the \textit{Euclid} Consortium, and the Danish National Research Foundation Cosmic Dawn Center (DNRF140); it is maintained by Stephane Rouberol.
The authors acknowledge the use of the Canadian Advanced Network for Astronomy Research (CANFAR) Science Platform operated by the Canadian Astronomy Data Centre (CADC) and the Digital Research Alliance of Canada (DRAC), with support from the National Research Council of Canada (NRC), the Canadian Space Agency (CSA), CANARIE, and the Canada Foundation for Innovation (CFI).
LWKG thanks the University of Edinburgh School of Physics and Astronomy for a postdoctoral Fellowship.
AG acknowledges the support of a grant from the Simons Foundation (Simons Investigator in Astrophysics, Award ID 620789).
HH is supported by a DFG Heisenberg grant (Hi 1495/5-1), the DFG Collaborative Research Center SFB1491, an ERC Consolidator Grant (No.\ 770935), and the DLR project 50QE2305.
MJH acknowledges support from NSERC through a Discovery Grant.
LVW acknowledges support from NSERC through a Discovery Grant.
We would like to thank our external blinding coordinator, Koen Kuijken.
We thank Douglas Scott for a thorough reading and detailed scientific and editorial comments.
This work made use of large language models (Claude, Anthropic; Codex, OpenAI) for code development, data analysis, and manuscript preparation.
All AI-assisted output was reviewed and validated by the lead author.
\end{acknowledgements}

\raggedbottom
\bibliographystyle{aa}
\bibliography{unions_bmodes}

\end{document}